\documentstyle[11pt,newpasp,twoside,epsf]{article}

\def\edcomment#1{\iffalse\marginpar{\raggedright\sl#1\/}\else\relax\fi}
\reversemarginpar

\def\mathfont#1{\ifmmode{#1}\else{$#1$}\fi} %for math font     
\def\lae{\mathrel{<\kern-1.0em\lower0.9ex\hbox{$\sim$}}}  
\def\gae{\mathrel{>\kern-1.0em\lower0.9ex\hbox{$\sim$}}}  
\def\msun{\ifmmode{\ {\rm M}_\odot}\else{$ {\rm M}_\odot$}\fi}  
\def\msunyr{\ifmmode{\msun \ {\rm yr}^{-1}}\else{$\msun \ {\rm yr}^{-1}$}\fi}  
\def\lae{\mathrel{<\kern-1.0em\lower0.9ex\hbox{$\sim$}}}  
\def\gae{\mathrel{>\kern-1.0em\lower0.9ex\hbox{$\sim$}}}

\def\msun{\ifmmode{\ {\rm M}_\odot}\else{$ {\rm M}_\odot$}\fi}  
\def\msunyr{\ifmmode{\msun \ {\rm yr}^{-1}}\else{$\msun \ {\rm yr}^{-1}$}\fi}

\input psfig.sty

\begin{document}

\title{Clusters of Galaxies at Sharp Focus: Cooling Flows and Radio
Sources Unveiled by Chandra}
\author{B. R. McNamara}
\affil{Dept. of Physics \& Astronomy, Ohio University, Clippinger Labs, 
Athens, OH 45701}

%\author{COAUTHOR NAME}
%\affil{COAUTHOR INSTITUTION, FULL ADDRESS, RUN TOGETHER}

\begin{abstract}
Chandra images of galaxy clusters have revealed a wealth 
of structure unseen by previous generations
of low resolution X-ray observatories.  In the cores of 
clusters, bright, irregular  X-ray emission is now routinely
seen within central dominant galaxies (CDGs) harboring
powerful radio sources. The radio sources 
are interacting with and are often displacing the hot gas
leaving cavities in their wakes.  The cavities rise buoyantly, 
transporting up to $10^{61}$ ergs of energy 
to the intracluster medium over the life of a cluster.
The dynamically
important magnetic fields observed in cooling flows may have
originated from these cavities.  The cooling rates found by Chandra 
and XMM-Newton are much smaller than previously reported, and they are
more in line with the levels of cold gas and 
star formation seen in other bandpasses.
I discuss the emerging correlation
between extended H$\alpha$ emission, star formation,
and bright X-ray emission at locations where the
cooling time of the keV gas approaches a few hundred million years. 
Data from a variety of disciplines suggest repeated, short duration
bursts of cooling and star formation are occurring, possibly 
induced and regulated in some instances by the central radio source. 
Finally, I discuss the interesting possibility that radio sources
quench cooling flows, while emphasizing the difficulties
with some proposed scenarios.

\end{abstract}

\section{Introduction}

In my oral presentation, I discussed four topics: the discovery
and interpretation of the large density discontinuities, or
so-called cluster ``cold fronts'',
Chandra observations of high redshift clusters, interactions
between radio sources and the hot intracluster medium (ICM),
and  cooling flows.
The first topic has been reviewed  by
Forman et al. (2001) and Sarazin (2001), to which I refer 
the interested reader.
The exciting topic of high resolution X-ray imaging of distant clusters 
is somewhat less mature, perhaps because it is photon-starved
and results are emerging slowly.  However, Chandra is beginning to
reveal the structure, temperatures, and masses of distant
clusters, giving us a new view of cluster formation and
evolution.  For further details, I refer the reader to
other discussions of high redshift galaxies 
(Fabian et al. 2001a, Holden et al. 2001, Stanford et al.
2001, Jeltema et al. 2001) while at the same time, 
apologizing to those whose papers I have missed.

I concentrate here on Chandra observations of the cores 
of relatively nearby $z<0.1$ cooling flow clusters (Fabian 1994).
I emphasize the bright X-ray systems where Chandra's superb spatial 
resolution can be used to full advantage in exploring the structure
of the ICM. These systems are particularly interesting as
they tend to harbor powerful radio sources.  Chandra 
observations of these systems are providing new insight 
into the life cycles of radio sources and the deposition of
energy and magnetic field into the ICM. In addition,
I highlight the significant contributions Chandra's
superb spatial resolution is making toward a solution to
the well-known cooling flow problem--the fate of the putatively
large cooling mass fluxes in clusters--which I discuss in
some detail.  This review updates and expands the briefer discussions
by McNamara et al. (2001a) and McNamara (2001). 

\section{Interactions Between Radio Sources and the keV Gas}

\begin{figure}
%\rule{5cm}{0.2mm}\hfill\rule{5cm}{0.2mm}
\hskip 0.2cm
\psfig{figure=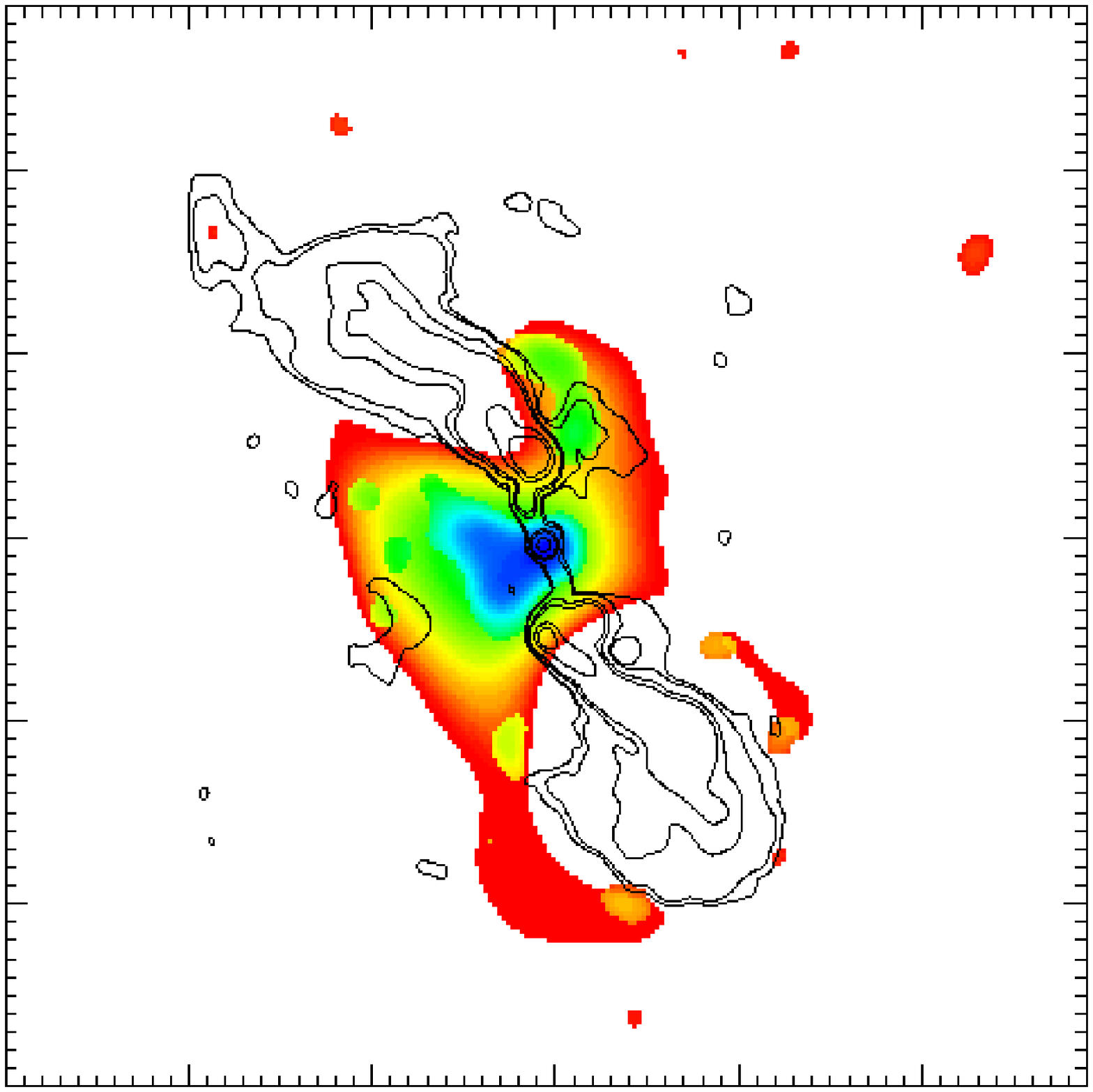,height=2.5in}
\psfig{figure=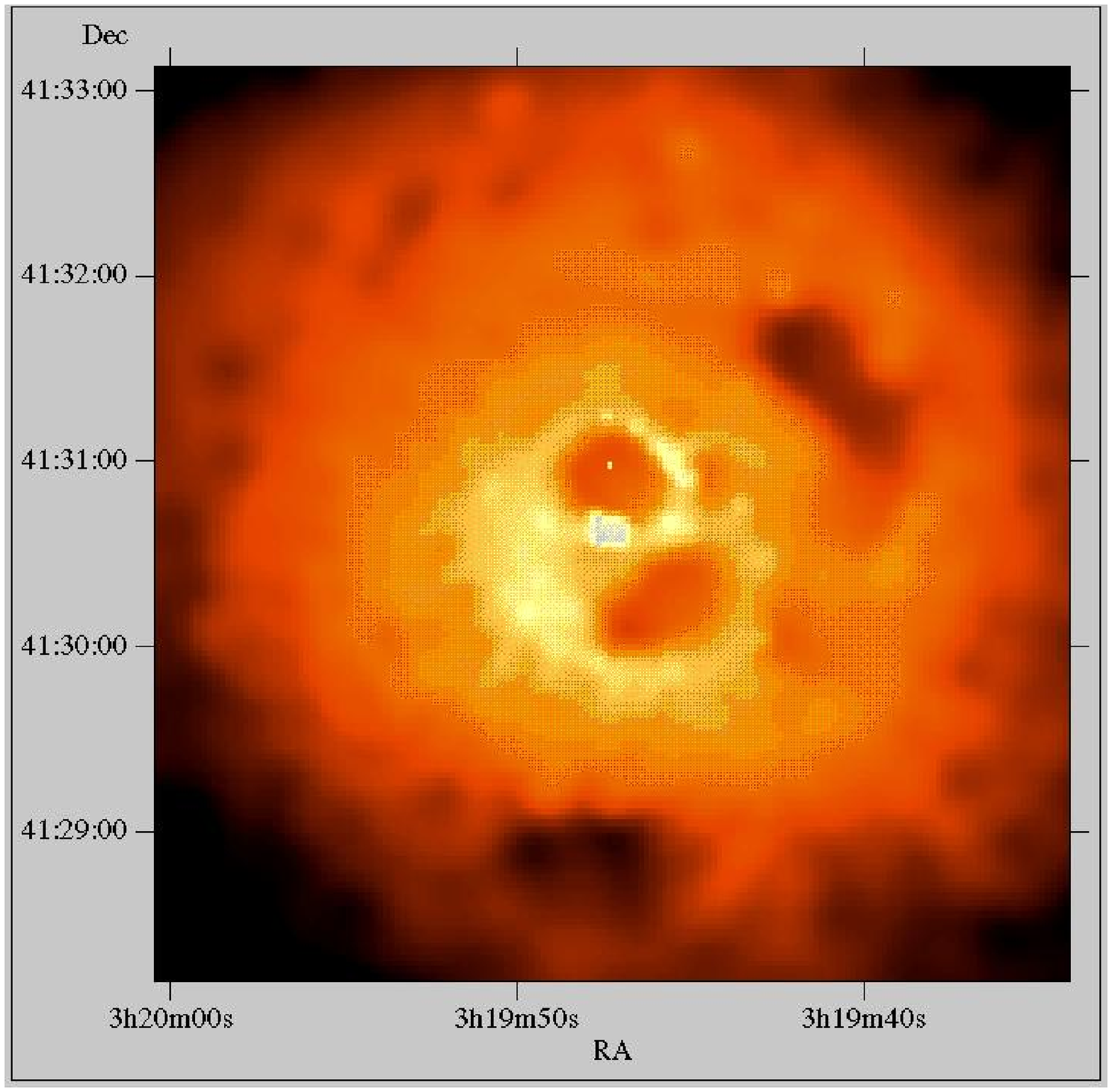,height=2.5in}
\caption{{\bf Left:} Chandra image of the central 
region of the Hydra A cluster (color). The X-ray image has been smoothed and 
filtered with wavelets in order to show the 
brightest emission. The 6 cm VLA Radio image of Hydra A
(contours) is superposed. See McNamara et al. (2000a) for details.
{\bf Right:} Smoothed Chandra image of the Perseus cluster centered
on NGC 1275 (Fabian et al. 2000a).  The inner cavities, seen just
north and south of center are filled with radio emission.  The
ghost cavities are seen roughly one arcmin north-west and south
of center. 
\label{fig:radish}}
\end{figure}

Images of cooling flow clusters taken in the past
decade with the {\it ROSAT} High Resolution Imager (HRI) have shown
X-ray structure surrounding the
radio sources in CDGs (Sarazin et al. 1995, Sarazin 1997, Rizza et al. 2000). 
The X-ray morphology varies, with 
some objects exhibiting enhanced emission at the location of the
radio source, while others show deficits.
However, in most instances, a detailed comparison
between the X-ray and radio morphologies was prohibited
by the roughly four arcsec resolution of the HRI.
By virtue of their proximity, the HRI images 
of NGC 1275 in the Perseus cluster (B\"ohringer et al. 1993) and
Cygnus A (Carilli, Perley, \& Harris 1994) clearly show
the radio lobes displacing the thermal gas surrounding them.  
Nevertheless, {\it ROSAT}'s limited spatial and spectral 
resolution did not permit a spectral analysis that would
lead to a detailed understanding of the nature of the interactions. 
Chandra has changed this situation dramatically.
Its exquisite, half arcsecond spatial resolution and moderate spectral
resolution is capable of probing both spatially and spectrally
interactions between radio sources and the surrounding plasma 
in a variety of systems including galaxies (Finoguenov \& Jones 2000, 
Jones et al. 2000), groups (Vrtilek et al. 2000), and
clusters.  I discuss here the recent cluster studies which
I suggest are leading to a
new understanding of these interactions and their
impact on the dynamics of the ambient medium in which they occur.

\subsection{Radio-Bright Cavities in the keV Gas}

The first Chandra images to show crisp signatures
of interactions between the keV gas
and powerful radio sources were those of the Hydra A and 
Perseus clusters (McNamara et al. 2000, Fabian et al. 2000a).
In both cases, the large depressions in the X-ray emission
coincident with the radio lobes and the bright emission surrounding
and between the lobes are strikingly shown in Figure 1.  
Since then many clusters have been reported with similar features, such
as Abell 2052 (Blanton et al. 2001), RBS 797 (Schindler et al. 2001),
MKW3s (Mazzotta et al. 2001), Abell 2597 (McNamara et al. 2001b),
Cygnus A (Wilson, Young, \& Shopbell 2000),
and 2A0335+096 in the Chandra archive.
As Figure 1 shows,  the radio lobes
appear to have displaced the keV gas surrounding them, leaving
cavities. Some of the questions to be answered
are how the cavities formed, how they are supported against collapse
(i.e., what is the equation of state within the
cavities?),  how long they survive, how frequently they occur, and
their long-term impact on the ICM.  I will address these topics here
to the varying degrees of detail permitted by the data available
at this time.

Using simulations of jets expanding 
into the ICM,  Clarke, Harris, \& Carilli (1997)
and Heinz, Reynolds, \& Begelman (1998) 
argued that the cavities seen in the {\it ROSAT} images of 
Perseus (B\"ohringer et al. 1993) and Cygnus A 
(Carilli, Perley, \& Harris 1994)
were caused by strong shock waves generated by relativistic jets.   
These models predicted that the X-ray emission from the rims 
surrounding the cavities
should be spectrally hard, and gas in the rim should have
higher entropy than the surrounding gas.  While these predictions
have yet to be tested in the case of Cygnus A, to our surprise
the Chandra images of Hydra A (McNamara et al. 2000, Nulsen et al. 2001),
Perseus (Fabian et al. 2000a), and Abell 2052 (Blanton et al. 2001) 
show soft emission from the rims, indicative of low entropy
gas. This implies that the radio lobes expanded gently into the
intracluster medium at roughly the sound speed 
in the keV gas (Reynolds, Heinz, \& Begelman 2001, David et al. 2001,
Nulsen et al. 2001).  The few cavities studied thus far 
do not appear to have formed
by shock waves associated with the radio sources, or they
are being viewed in the late stages of their development.

\begin{figure}
%\rule{5cm}{0.2mm}\hfill\rule{5cm}{0.2mm}
\hskip 0.4cm
\psfig{figure=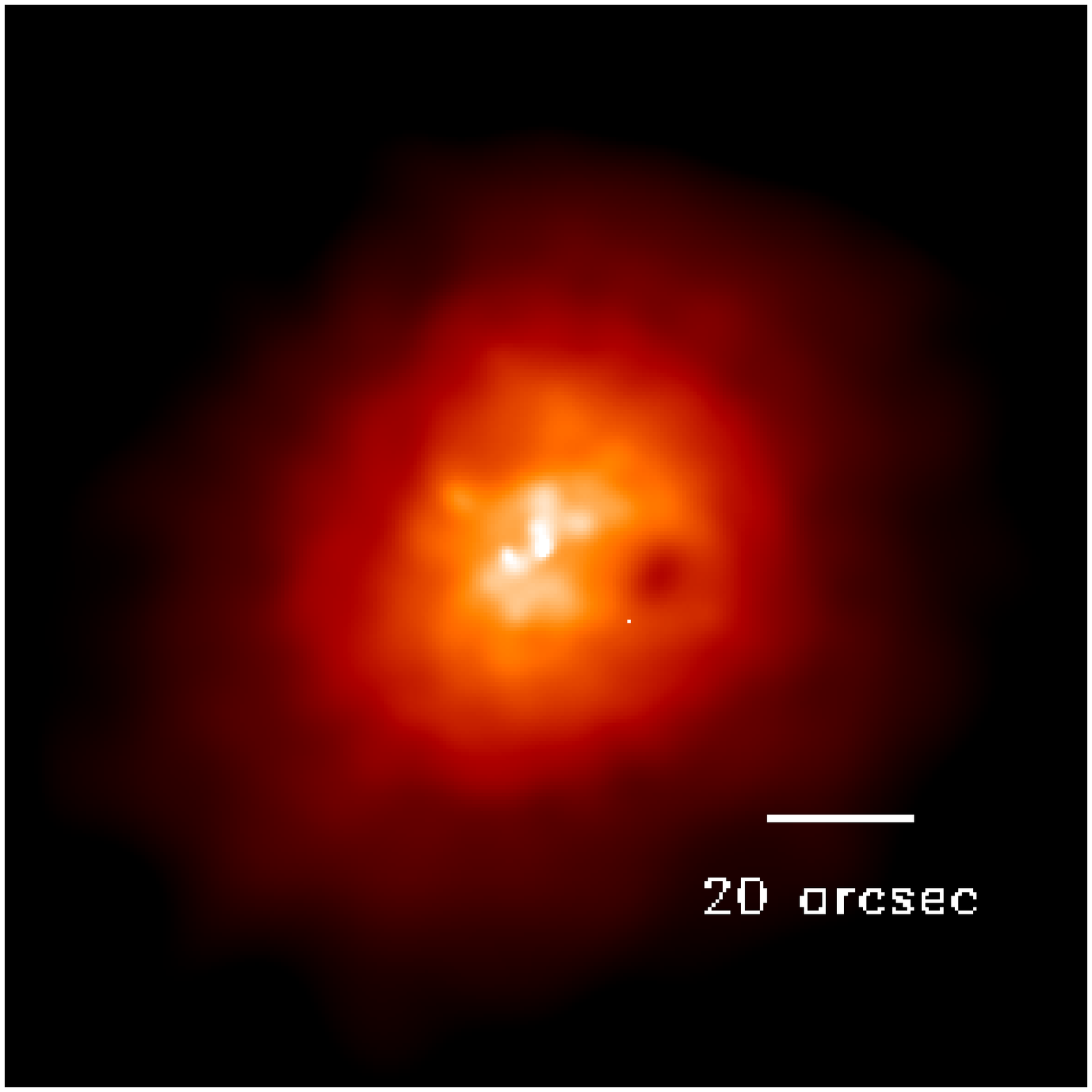,height=2.5in,angle=90}
\psfig{figure=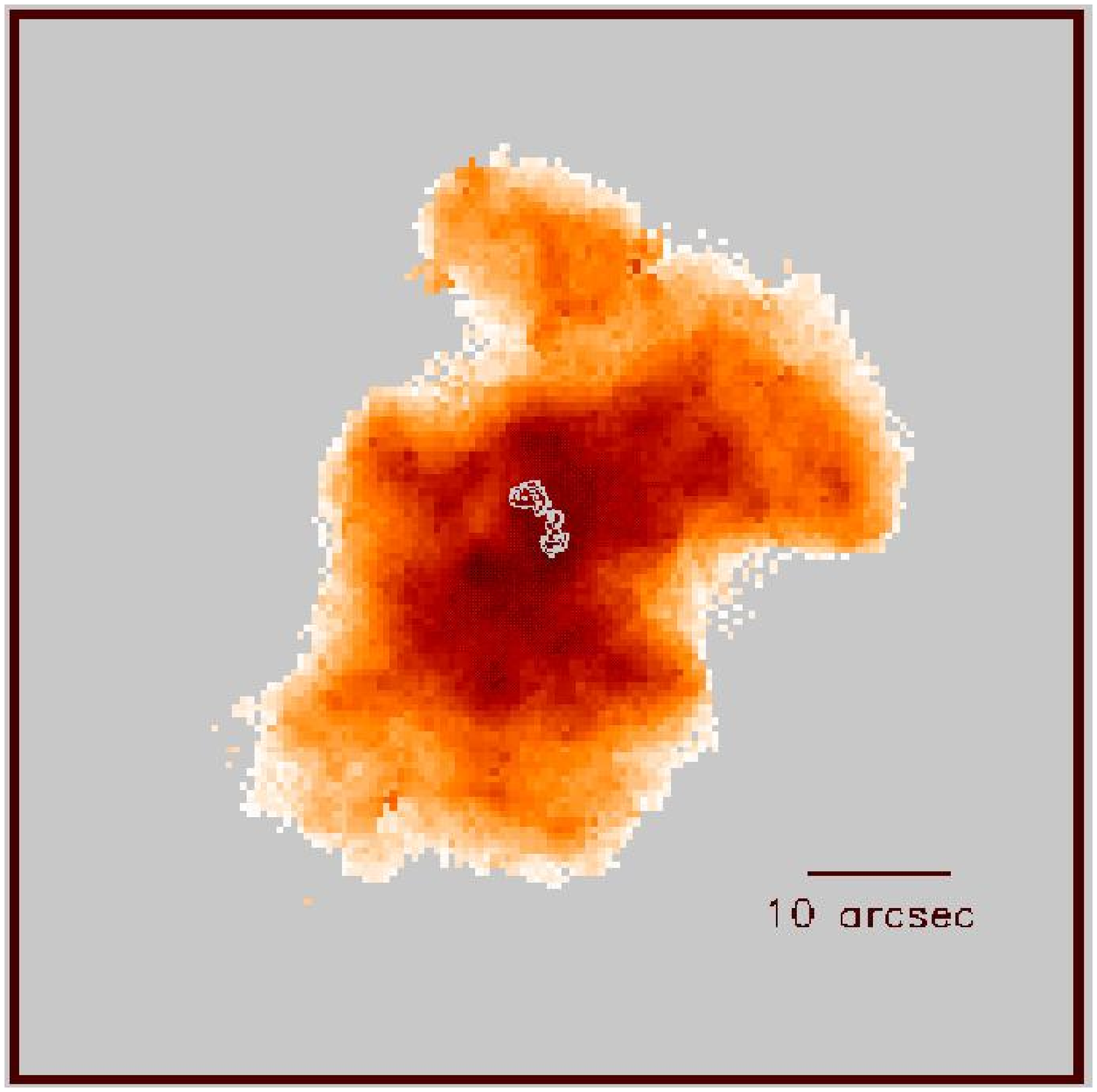,height=2.5in,angle=180}
%\rule{5cm}{0.2mm}\hfill\rule{5cm}{0.2mm}
\caption{{\bf Left:} Smoothed Chandra image of Abell 2597. Note the
irregular surface brightness in the inner 40 arcsec. The surface
brightness depressions associated with the ghost cavities are seen
18 arcsec southwest and northeast of center. {\bf Right:}  Expanded view
of the of the central region of Abell 2597 after subtracting a smooth 
background cluster model.  The 8.44 GHz radio contours are superposed.
The ghost cavities are the indentations in the emission. 
North is at top; east is to the left in both figures (McNamara et al. 2001b).
\label{fig:rad}}
\end{figure}

The source of pressure within the cavities is not well
understood.  In Hydra A and Abell 2052, the external gas pressure exceeds the
internal equipartition magnetic field and cosmic ray pressure 
by roughly a factor of ten (McNamara et al. 2000, Blanton 
et al. 2001).  If the equipartition pressures are valid, 
the cavities should quickly collapse. However, their high frequency of
detection, and the existence of the older, radio-quiet ghost cavities,
implies that they persist on timescales longer than their
sound crossing (collapse) timescales.  The cavities should then
be roughly in pressure balance with the surrounding gas. 
This would require an additional source of  
internal pressure support such as a dilute thermal gas,
or an unseen population of energetic particles.    

In Hydra A and Abell 2052, the
data are consistent with the cavities being devoid of thermal
gas. However, the uncertainties of the measurements
do not preclude this.  As a counter example, 
Mazzotta et al. (2001) have reported
a hot thermal bubble in  MKW3s. 
In either case, the gas density within the cavities is clearly less than their
surroundings, so they should behave
like bubbles in water and rise buoyantly in the intracluster 
medium (McNamara et al. 2000, Churazov et al. 2001).  
In this instance, the timescale for cavities to reach their
heights in cluster atmospheres is determined by their
buoyancy speed, which is of the same order as the Keplerian
velocity at their true distance from the center of the 
cluster (Churazov et al. 2001).  This timescale
provides crude ages for the cavities of
$\sim 10^{7.5}-10^8$ yr, which exceeds their sound
collapse timescales.

\subsection{Radio-Faint ``Ghost'' Cavities in the keV Gas}

In addition to radio-filled cavities, holes in the cluster
X-ray emission that lack 
bright radio emission above $\sim 1$ GHz have been discovered in
the Perseus and Abell 2597 clusters.  These ``ghost'' 
cavities are seen in Perseus beyond its inner 
radio-filled cavities (Fabian et al. 2000a and Figure 1) and beyond
the powerful radio source in the nucleus of Abell 2597's CDG
(McNamara et al. 2001b and Figure 2).  In Perseus, flattened ghost cavities
are seen roughly one arcmin to the northwest and south of
the cluster center (Figure 1).  The inner
cavities are filled with bright radio emission (Fabian et al. 2000a), 
although the radio emission is not shown in Figure 1. 

Likewise, Abell 2597's ghost cavities are located symmetrically
about the inner radio source roughly 20 arcsec
to the northeast and southwest of the CDG's nucleus.  The cavity
to the southwest is most prominent in the left panel of Figure 2.
In the right panel of Figure 2, I have displayed the cavities
with respect to the compact central radio source
after subtracting the background cluster emission and expanding
the image to highlight the central structure.  The ghost cavities
clearly do not coincide with the central radio source.   

Based on the properties of radio-bright cavities, 
it is reasonable to suppose that the ghost cavities were produced 
in radio episodes that predated the current ones.  
One might then expect to detect faint radio emission from 
the dying lobes of the causal radio event.
Indeed, faint spurs of radio emission have been detected
near Perseus's ghost cavities at 74 MHz (Blundell et al. 2000,
Fabian et al. 2000a).  Likewise, faint radio emission at 1.4 GHz
was detected at the level of a few mJy toward Abell 2597's
ghost cavities with the VLA (McNamara et al. 2001b).  These 
radio detections are consistent with the hypothesis that
the ghost cavities were created in earlier radio events
that occurred between 50--100 Myr ago. The
energy required to inflate Abell 2597's ghost cavities
implies that the radio source
was an order of magnitude more powerful in the past than the one
seen today, which would have been as powerful as Hydra A.  These bubbles
of magnetic field and cosmic rays are presumably lifted
into the outer regions of clusters where, rejuvenated,
they may contribute
to the formation of radio halos (Kempner \& Sarazin 2001, 
En\ss lin \& Gopal-Krishna 2001).  

\subsection{Repeated Radio Outbursts Magnetize the keV Gas}

That nearly 70\% of CDGs in cooling flows are radio-bright (Burns 1990)
implies that radio sources live longer than about 1 Gyr,
or they recur with high frequency.  Our interpretation of the cavities
implies the latter, with a recurrence approximately every 100 Myr
(McNamara et al 2001b). This would have significant implications for
understanding energy feedback from the radio source
to the ICM (Tucker \& David 1997, Soker et al. 2001, Churazov et al. 2002).
A lower limit to the energy expended during cavity
formation is given by the  $PdV$ work done on the
surrounding gas.  For Hydra A, the total energy is $\sim 10^{59}$
erg (McNamara et al. 2000), and energies of the same order
are found for Abell 2052 (Blanton et al. 2001) and 
Abell 2597 (McNamara et al. 2001b).  

Assuming that CDGs produce between $10-100$ bubbles over their lifetimes, 
each with an energy of $\sim 10^{58-59}~{\rm erg}$, 
they would deposit $\gae 10^{59-61}$ erg 
into the ICM in the form of magnetic field, cosmic rays, and 
heat.  This energy would be comparable to the
the total thermal energy of the X-ray-emitting plasma in the
inner regions of some clusters. Clusters are
magnetized  (Clarke, Kronberg, \& B\"ohringer 2001, Kronberg et al. 2001), 
and these bubbles may be vessels 
that transport magnetic field from giant, central black 
holes to the ICM. If a significant fraction this energy were
deposited as magnetic field in the inner 100 kpc of clusters, the 
implied field strengths of $\sim 5-50 \mu$G  would be consistent with
the field strengths derived from Faraday rotation measures
observed in the cores of cooling flow clusters (Ge \& Owen 1993).

\section{The Cooling Flow Problem in the Chandra--XMM-Newton Era} 

Galaxy clusters often harbor 
bright cusps of X-ray emission in
their central $\sim 100$ kpc.  These cusps are  associated with regions of 
dense gas with radiative
cooling times less than a few Gyr (Fabian 1994).  
Absent a significant source of heat,
the gas will cool to low temperatures and accrete onto
the cluster's CDG where it will presumably accumulate 
in atomic and molecular clouds and form stars.  
Indeed, the likelihood that a CDG
has detectable levels of cold gas and star formation increases
dramatically with X-ray cooling rate (McNamara 1997, Cardiel et al. 1998,
Allen 1995, Crawford \& Fabian 1993, Crawford et al. 1999, Edge 2001).
This is shown in Figure 3, where I plot the trend for larger blueward
$U-B$ colors associated with star formation in CDGs with increasing
mass cooling rate.  The cooling rates were estimated 
primarily with {\it ROSAT} data. The correlation suggests
a strong connection between cooling flows and star formation.
The problem with this interpretation rests with the large
discrepancies between the star formation rates implied by
the $U-B$ colors and the X-ray cooling rates.

\begin{figure}
%\rule{5cm}{0.2mm}\hfill\rule{5cm}{0.2mm}
\hskip 1.0cm
\psfig{figure=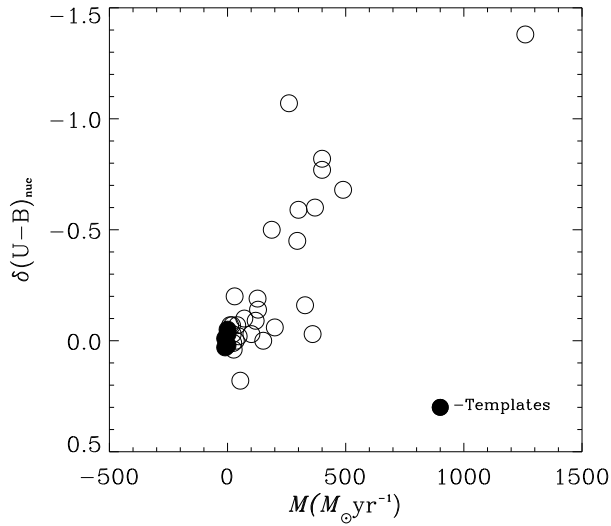,height=3.0in}
%\rule{5cm}{0.2mm}\hfill\rule{5cm}{0.2mm}
\caption{Correlation between central $U-B$ continuum
color excess and total X-ray cooling rate.  The filled points represent
non-accreting galaxies with normal colors.
\label{fig:rad}}
\end{figure}

While the cooling rates plotted are typically hundreds of solar 
masses per year,
the star formation rates are typically several to several tens of
solar masses per year, depending on the star formation history 
of the system in question. The systems with the
largest star formation rates, which approach $\sim 100 \msunyr$,
appear to have experienced
bursts or episodes of star formation lasting $\lae 100$ Myr or so
(McNamara 1997).  Therefore, unless cooling is likewise episodic,
cold gas would accumulate to the observed levels
in tens of Myr, much less than the
probable $1-10$ Gyr ages of cooling flows. This uncomfortable 
situation has lead to the
view that either the matter is accumulating in a dark or
otherwise unusual physical state,  or the cooling rates have been 
substantially overestimated.  The Chandra and XMM-Newton data
are showing that the cooling rates were indeed 
substantially overestimated.  Nevertheless, the consensus of
data from a variety of disciplines suggests that
cooling is indeed occurring.

\subsection{The Short Cooling Time of the keV Gas}

Chandra has the unprecedented capability to measure
the state of the keV gas in clusters
on scales of a few arcsec.  This corresponds to a few kpc for 
clusters within $z\lae 0.1$, which is comparable to the 
sizes of the radio sources and sites of 
star formation.  With this unique capability, 
Chandra is applying a critical test of
the cooling flow paradigm by providing for a direct comparison 
between local cooling rates and star formation rates, 
and searches for local heat sources that could
reduce the cooling rates.  
\begin{figure}
%\rule{5cm}{0.2mm}\hfill\rule{5cm}{0.2mm}
\hskip 1.5cm
\psfig{figure=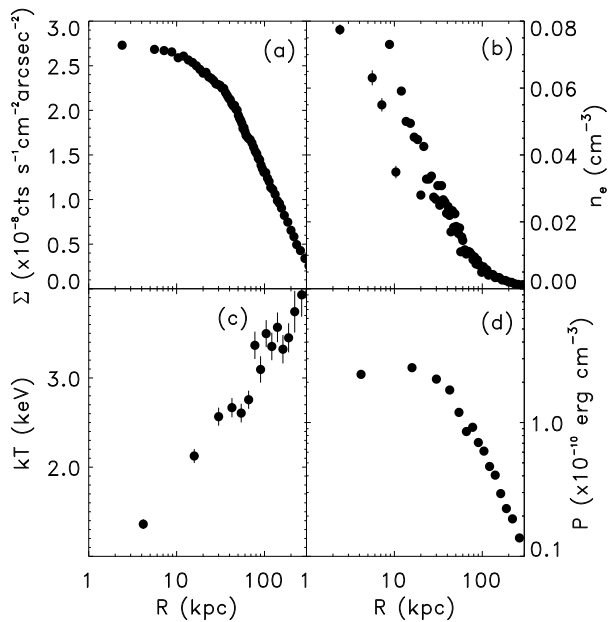,height=3.5in}
%\rule{5cm}{0.2mm}\hfill\rule{5cm}{0.2mm}
\caption{ Radial variation of (a) surface brightness, 
(b) electron density,
(c) temperature, (d) pressure in Abell 2597 (McNamara et al. 2001b). 
\label{fig:rad}}
\end{figure}
Chandra has confirmed the signature
decline in gas temperature and the steeply rising
density and pressure in the inner few tens of kpc of cooling
flows (McNamara et al. 2000,
David et al. 2001, Blanton et al. 2001, McNamara et al. 2001b).
A typical example, Abell 2597, is shown here in Figure 4. 
The gas temperature declines from $\simeq 3.5$ keV at 100 kpc
to $\simeq 1.3$ keV in the inner few kpc.  At the same time,
the density and
pressure increase dramatically, reaching
values of $0.08~{\rm cm}^{-3}$ and $1.1\times 10^{-10}~{\rm erg~cm}^{-3}$
in the central few kpc.  The radiative cooling time of the gas 
there is only a few hundred Myr.  
No other X-ray observatory can make these measurements on such a fine scale. 

\subsection{Correlated Sites of Cooler KeV Gas \& Star Formation} 

Several Chandra studies have shown a great 
deal of X-ray structure associated with high density, cooler gas
in the central few tens of kpc of clusters.  
Much of this structure, seen for example in Figures 1 \& 2,
is associated with the radio sources, nebular
emission, and sites of star formation.
For instance, a bright, flattened X-ray structure is associated
with the disk of star formation and nebular emission
in Hydra A (McNamara et al. 2000).  A similar spatial correlation
is seen in Abell 2597, where bright knots of X-ray emission
accompany the regions of ongoing star formation, nebular emission,
and molecular gas (Heckman et al. 1989, 
McNamara \& O'Connell 1993, Koekemoer et al. 
1999, Voit \& Donahue 1997, Donahue et al. 2000, Baker \& Jaffe 2001,
private communication).
In both Hydra A and Abell 2597, the star formation regions
and H$\alpha$ emission are seen
where the cooling time of the keV gas is shortest $\sim 300-600$ Myr.

\begin{figure}
%\rule{5cm}{0.2mm}\hfill\rule{5cm}{0.2mm}
\hskip 0.0cm
\psfig{figure=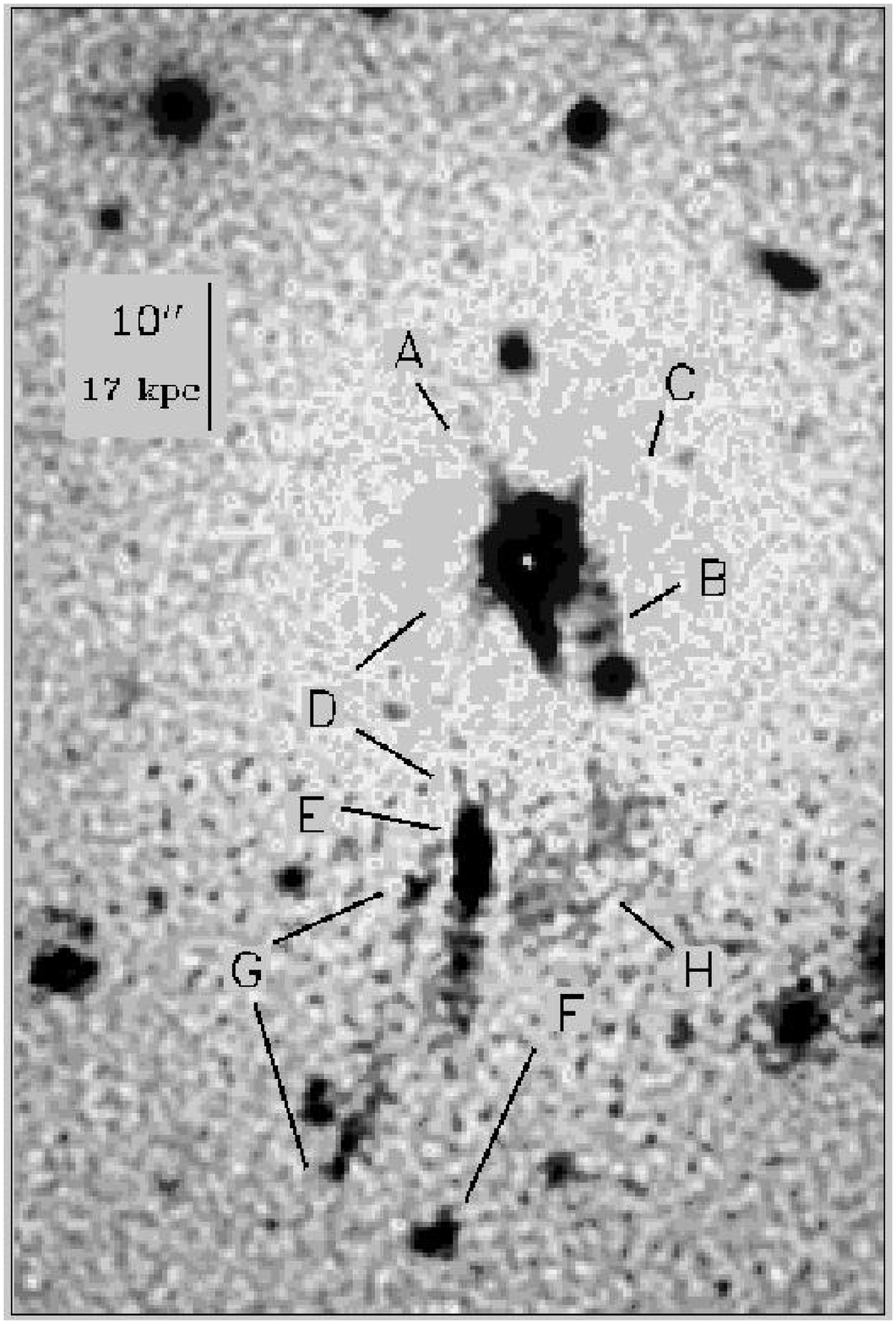,height=3.0in}
\psfig{figure=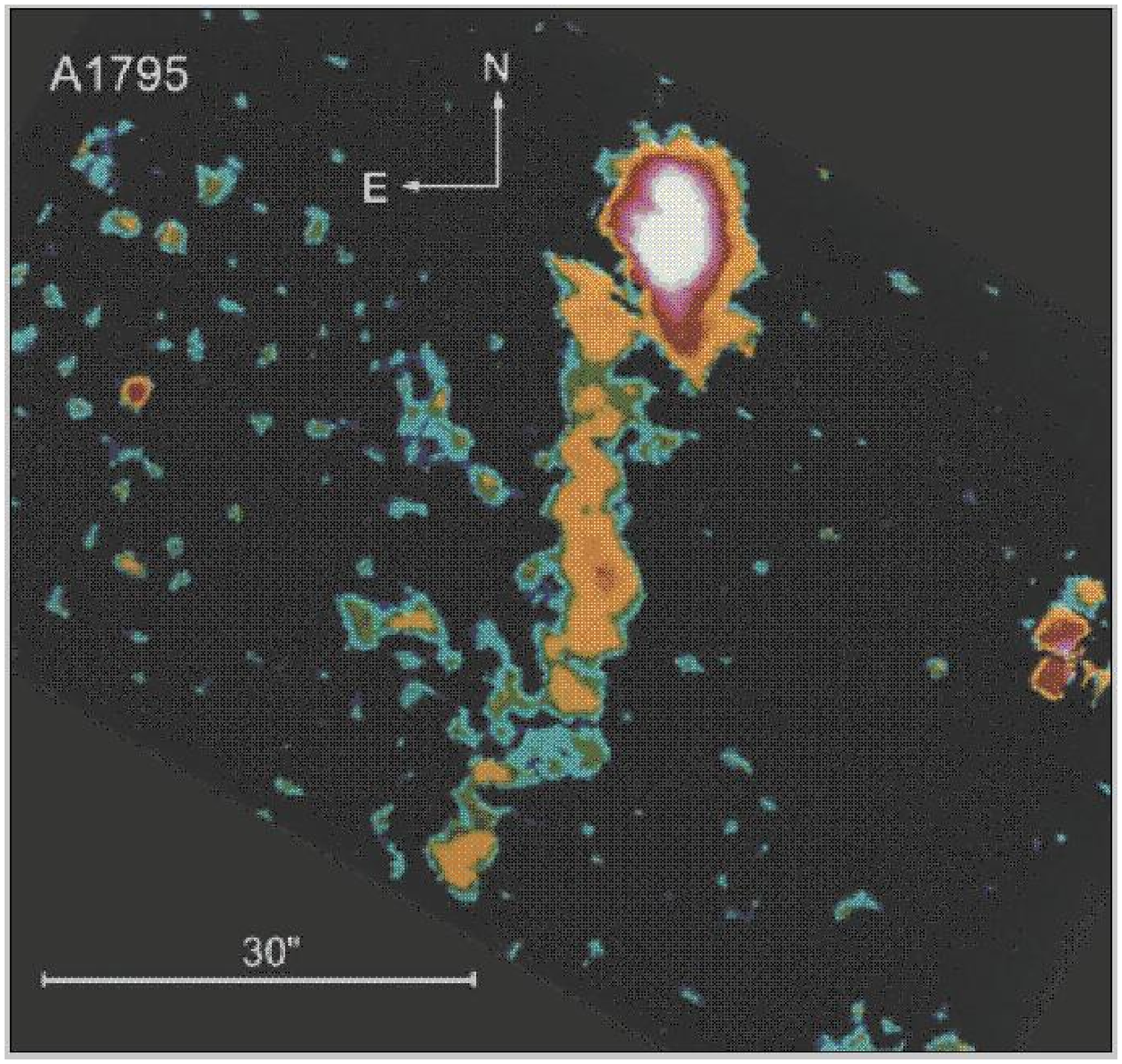,height=3.0in}
\psfig{figure=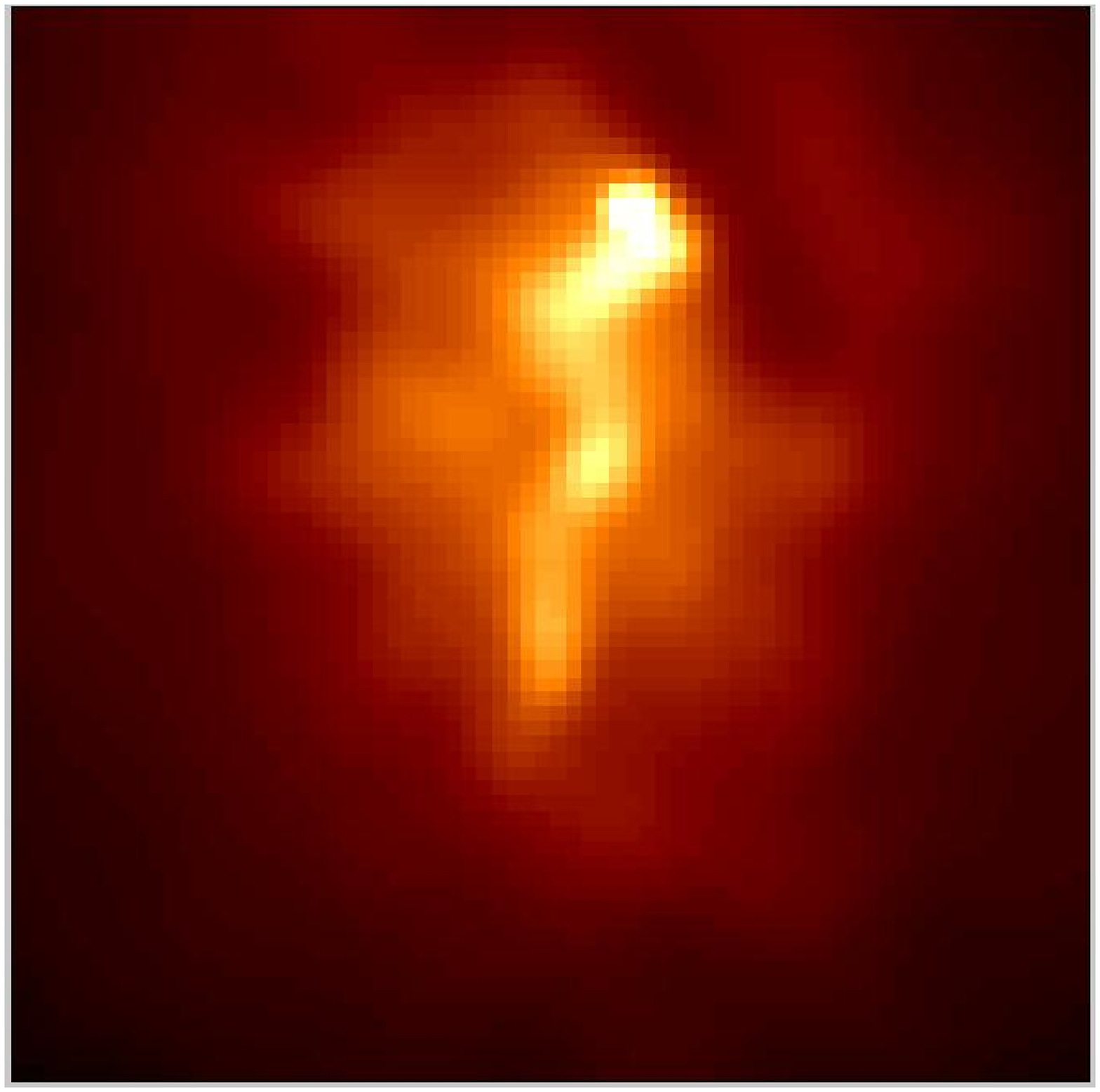,height=2.0in}
%\rule{5cm}{0.2mm}\hfill\rule{5cm}{0.2mm}
\caption{{\bf Upper Left:} Star formation map (McNamara et al. 1996),
{\bf Upper Right:} H$\alpha$ map (Cowie et al. 1983), and {\bf Lower Left:}
Chandra X-ray map of the center of the Abell 1795 cluster (Fabian et al. 2001b).  The trail of
star formation, indicated by lettering, is traced using a processed,
deep U-band image (McNamara et al. 1996).  The U-band, H$\alpha$,
and X-ray trails are clearly correlated.  
\label{fig:rad}}
\end{figure}

The most striking example of this correlation is seen in Abell
1795.  The Chandra image (Fabian et al. 2001b) shows 
a bright filament of gas near the nucleus of
the CDG that extends southward between 50--70 kpc.
Trails of $U$-band, young-star continuum (McNamara et al. 1996) and 
H$\alpha$ emission (Cowie et al. 1983)
roughly correspond to the bright X-ray filament.  
I have reproduced these images
here in Figure 5.  Although it is not clear how these features came 
into being, the possibilities include a cooling
filament or ``cooling wake'' of gas that formed as the CDG sloshed about
in the cluster potential (cf., Markevitch, Vikhlinin, \& Mazzotta 2001),
or the remains of cooler gas that arrived in the cluster core 
by way of a merger with a group of galaxies (McNamara et al. 1996).  
In any case, these
correlations clearly establish a relationship between the coolest 
X-ray-emitting gas and star formation, as is suggested by
the statistical correlation between cooling rate and color excess
shown in Figure 3.

\subsection{Atomic \& Molecular Gas of Varying Temperature}

I wish to emphasize that the evidence for gas with 
a range of temperature, as would be expected in a cooling flow, is
not restricted to keV and recombination (H$\alpha$) temperatures alone.
Large reservoirs of atomic and molecular gas with a variety of
temperatures are seen in cooling flows.  
H I has been detected in absorption against
the radio continua of several CDGs in large cooling flows (e.g., O'Dea,
Baum, \& Galimore 1994).  The kinetic temperature of the
atomic hydrogen is probably between $100-1000$ K. 
In addition, molecular gas has been detected in
the $2.1\mu$ H$_2$ feature in essentially all of the 
large cooling flows with strong nebular line emission (Elston \&
Maloney 1994, Jaffe \& Bremer 1997, Falcke et al. 1998,
Donahue et al. 2000).  This
infrared feature is tracing molecular gas at a temperature
of $\sim 2000$ K. 
Furthermore, carbon monoxide emission, which was first detected in 
Perseus (Lazareff et al. 1989), has now been detected in many
cooling flow clusters by Edge (2001).  In several cases, 
emission is seen in multiple line transitions which seem to
trace  warm molecular gas at $\simeq 20-40$ K.  The amounts of
gas can exceed $10^{10}\msun$ in some cases, bearing in mind
uncertainties in the mass estimates of factors of several.  
These properties are atypical of CDGs in clusters 
without the X-ray signatures of cooling flows, although sensitive
observations of a proper control sample should be pursued to confirm this.   
In any case, these observations demonstrate that large reservoirs
of gas ranging between the virial temperature and molecular cloud temperatures
are indeed present in the central few tens of kpc of cooling flows.

\subsection{Reduced Cooling Rates from Chandra and XMM-Newton}

Cooling rates have been now been estimated using a variety of methods
and instruments on Chandra and XMM-Newton for several
cooling flow clusters (Fabian et al. 2000b,
McNamara et al. 2000, Peterson et al. 2001, David et al. 2001, 
Molendi \& Pizzolato 2001, B\"ohringer et al. 2001).  
The consensus reached in these
studies of an admittedly limited number of objects is
that the spectroscopically-derived cooling rates are factors of 5--10
less than earlier {\it ROSAT} and {\it Einstein} values.

The most compelling evidence for reduced cooling
is found with the XMM-Newton and Chandra high resolution 
grating spectra for the putatively massive, $\dot M\sim 2300 \msunyr$ 
cooling flow in Abell 1835 (Allen et al. 1996). 
In contrast, the grating spectra give upper 
limits of $\dot M \lae 200-300\msunyr$,
based primarily on Fe XVII emission line fluxes
(Peterson et al. 2001, Wise 2001, private communication).  
These measurements imply
that either most of the cooling gas is maintained
above $\sim 2$ keV,
or that gas is cooling without an obvious spectroscopic signature  
(Fabian et al 2000b).  Moderate resolution CCD spectroscopy
from Chandra and XMM-Newton also give systematically lower cooling rates.

The only remaining evidence for massive cooling
is found using {morphological} cooling rates, calculated
essentially by dividing the central gas mass
by the cooling time.  This method gives $\dot M$s that are
roughly consistent with results from earlier X-ray missions.
However, the morphological $\dot M$ vastly exceed
the spectroscopic $\dot M$ in Hydra A, for example (David et al. 2001).
The inconsistency between these methods
suggests that either the cooling gas is being reheated
or maintained at keV temperatures by some heating 
process (David et al. 2001, Fabian et al. 2000b, Narayan \& Medvedev 2001), 
or our cooling models are wrong.

\subsection{A Stellar Repository for Cooling Gas in Abell 2597 \& Hydra A}

The strongest spectroscopic evidence for cooling
gas in the Hydra A cluster is found in the central 30 or so 
kpc (David et al. 2001, McNamara et al. 2000) near 
a large circumnuclear disk of young stars and gas (McNamara
1995, Hansen et al. 1995, Melnick et al 1997).  
The spectroscopic cooling rate derived from spectral imaging 
is $\dot M\simeq 35 \msunyr$  within a 70 kpc radius.  Beyond roughly
70 kpc, the Chandra data are consistent with single temperature
thermal models without cooling, while earlier studies
reported cooling rates of
$\dot M \sim 300-600 \msunyr$ (David et al. 1990, 
Peres et al. 1998, but see Ikebe et al. 1997).  
The mass of young stars in the disk derived from 
rotation curve and luminosity measurements is
$10^{8\to 9} \msun$.  This mass is consistent
with the present rate of cooling throughout the limited volume of
the disk over the age of the cluster.  
However, if all $35 \msunyr$ of material
cooling within a 70 kpc radius is accreting onto the disk,
it would double its mass in $\sim 30$ Myr. 

Similarly, a preliminary analysis of a
Chandra image of Abell 2597 gives $\dot M\simeq 60 \msunyr$
(McNamara et al. 2002),
compared to the $ROSAT$ PSPC value of $\dot M\simeq 280-350 \msunyr$ 
(Sarazin \& McNamara 1997, Peres et al. 1998).  
The CDG in Abell 2597 likewise 
is experiencing  vigorous star formation with a total
mass of $\sim 10^{8\to 9}\msun$ 
(McNamara \& O'Connell 1993, Koekemoer et al. 1999).  
Cooling at the above rate would be capable of fueling star formation
for a few tens of Myr or so depending on the star formation history.
The upshot here is that the {\it local} cooling rates surrounding
the star formation regions and the
star formation rates seem to be converging.  However, the
state of the hot gas in cluster cores is much more complex
than we might have hoped. Until these complexities are
sorted out, the connection between cooling and star formation
will remain uncertain.
A lingering issue with the X-ray analysis is that
while cooling flow models generally reproduce the observed
spectra of the inner most regions of clusters quite well,
we and others (Molendi \& Pizzolato 2001) find that single 
temperature models often work equally well.  
There is ample evidence for keV gas with short cooling
times, but there is little evidence for multiphase, cooling gas. 

\subsection{Episodic Cooling, Star Formation, and Feedback}

The properties of Abell 2597 and Hydra A, 
if representative of other cooling flows,
suggest that cooling and star formation occur episodically,
rather than in a long duration flow as was previously thought.
Interestingly, it has been known
for some time that star formation in cooling flows probably
occurs primarily in bursts, and that these bursts are
sometimes associated with the radio source
(McNamara \& O'Connell 1993, Allen 1995, McNamara 1997, Cardiel et al. 1998).
Furthermore, the star formation episodes occur on timescales that
are less than or on the order of both the $\sim 100$ Myr timescales for radio outbursts inferred
from the ghost cavity properties, and the cooling time of the gas
in the vicinity of the star formation regions.  
The similarity of these timescales is at least 
consistent with the notion that feedback between gas 
accretion and the radio source
is in some cases triggering star formation, while at the same time,
regulating the degree of cooling.

One of the most appealing models to explain the reduced cooling rates
is heating by the mechanical power
of radio sources (Soker et al. 2001, Churazov et al. 2002).
While there is some evidence for heating as seen in
Hydra A's flat central entropy profile (David et al. 2001),
the gas surrounding the radio source is {\it cooler}, not hotter, 
than the ambient gas (McNamara et al. 2000, Fabian 2000a, 
Nulsen et al. 2001).  This fact is seriously
troubling for direct heating scenarios.

\begin{figure}
%\rule{5cm}{0.2mm}\hfill\rule{5cm}{0.2mm}
\hskip 1.5cm
\psfig{figure=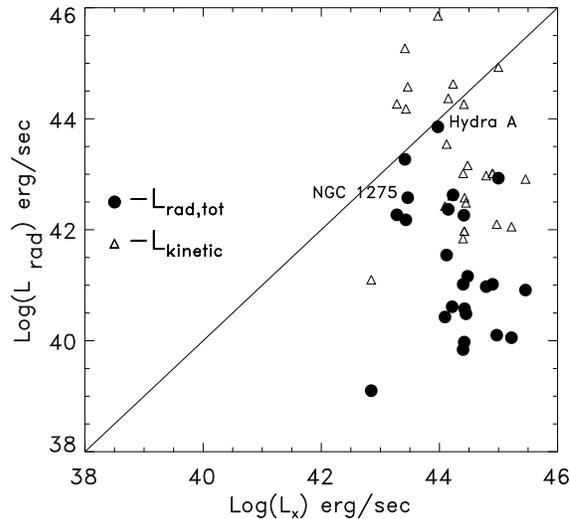,height=3.0in}
\caption{Correlation between central X-ray luminosity,
total radio luminosity (filled circles), and kinetic luminosity
(triangles, assuming $L_{\rm kinetic}=100\times L_{\rm rad, tot}$).
\label{fig:rad}}
\end{figure}

In Figure 6, I plot radio luminosity
against central X-ray luminosity for a small sample of cooling flows.
The filled points represent the observed radio luminosity, while
the triangles are the same points multiplied by $100$ 
to represent the possible total mechanical energy available
to work on the surrounding medium. The line indicates the
threshold where the instantaneous radio luminosity matches
the cooling luminosity, or the minimum
energy required to quench a cooling flow over the short duration
of the radio source. 

Figure 6 shows that Hydra A and a few other sources
are certainly powerful enough to significantly
reduce or quench a cooling flow for many tens of Myr or so.  
However, most objects do not have enough
power to quench cooling, and those that do must efficiently
couple their mechanical power to the cooling gas.  
How the radio power would be coupled to the cooling gas is poorly
understood.  Several authors (David et al. 2001, Nulsen et al. 2001,
Quilis, Bower \& Balogh 2001, Br\"uggen et al. 2002, Churazov et al. 2002) have
proposed scenarios where convective currents driven by Hydra A's
radio source may be removing cooling material out of 
the cluster's core where it will expand, mix with
ambient gas, and cool less efficiently. 
This process can assist in reducing the deposition
of cooled gas without directly introducing heat. 
However, it is not clear whether these scenarios 
are consistent with the pronounced metallicity gradient in the cluster's core 
(David et al. 2001,  Fukazawa et al. 2000).
Convection should remove a metallicity
gradient, unless the excess metals can be replenished rapidly
by ongoing star formation.

I should caution that the {\it ROSAT} data plotted in Figure 6 were
taken heterogeneously from the literature,
and the spatial resolution is poorly matched to the radio sources.
Certainly this correlation needs to be revisited with Chandra.
But it serves to illustrate the point that although
radio sources clearly have a major impact on the cores
of clusters, their ability to quench cooling over cluster
ages has not, in my view, been demonstrated.  
Other agents such as heat conduction from the hot 
outer layers of clusters ought, yet again, to be reconsidered 
(Narayan \& Medvedev 2001) in models aiming to reduce the level of
cooling in cooling flows.

\section*{Acknowledgments}

I would like to acknowledge the colleagues with whom I have worked
closely in the last few years, Michael Wise, Paul Nulsen, 
Craig Sarazin, Larry David, and Chris Carilli, 
for their major contributions to the 
work discussed here.  This research is supported by generous grants
from NASA, the Chandra X-ray Center, the Space Telescope Science
Institute, and the Department of Energy, including LTSA grant NAG5-11025
and Chandra General Observer and Archival Research Awards GO0-1078A, 
AR2300-7X, and GO1-2139X.  I would like to give special thanks
to Eric Schlegel for organizing this meeting, and to the Scientific
Organizing Committee for inviting me to attend.

\end{document}